\documentclass[amsmath,amssymb, 12pt,a4paper, tightenlines]{revtex4-1}

\usepackage[usenames,dvipsnames,svgnames,table]{xcolor}

\usepackage{graphicx}
\usepackage{epstopdf}
\usepackage{amssymb,amsfonts,amsmath}
\usepackage{siunitx}
\usepackage{bm}

\usepackage{color}
\definecolor{blue}{HTML}{377EB8}
\definecolor{red}{HTML}{E31A1C}

\usepackage[usenames,dvipsnames,svgnames,table]{xcolor}

\usepackage{natbib}
\usepackage{amssymb,amsfonts,amsmath, booktabs, cancel, csquotes, datatool, helvet, mathpazo, multirow, listings, pgfplots, wasysym}
\usepackage{bm}
\usepackage{placeins}

\AppendGraphicsExtensions{.tif}

\usepackage{color}
\definecolor{blue}{HTML}{08519C}
\definecolor{red}{HTML}{99000D}
\definecolor{yellow}{HTML}{FFCC00}

\begin{document}

\title{The cost of swimming in generalized Newtonian fluids: Experiments with \textit{C. elegans}}

\author{David~A.~Gagnon}
\author{Paulo~E.~Arratia}
\email{parratia@seas.upenn.edu}
\affiliation{Department of Mechanical Engineering and Applied Mechanics, University~of~Pennsylvania, Philadelphia,~PA 19104}


\date{15 May 2016}

\begin{abstract}
Numerous natural processes are contingent on microorganisms' ability to swim through fluids with non-Newtonian rheology. Here, we use the model organism \textit{Caenorhabditis elegans} and tracking methods to experimentally investigate the dynamics of undulatory swimming in shear-thinning fluids. Theory and simulation have proposed that the cost of swimming, or mechanical power, should be lower in a shear-thinning fluid compared to a Newtonian fluid of the same zero-shear viscosity. We aim to provide an experimental investigation into the cost of swimming in a shear-thinning fluid from \textit{(i)} an estimate of the mechanical power of the swimmer and \textit{(ii)} the viscous dissipation rate of the flow field, which should yield equivalent results for a self-propelled low Reynolds number swimmer. We find the cost of swimming in shear-thinning fluids is less than or equal to the cost of swimming in Newtonian fluids of the same zero-shear viscosity; furthermore, the cost of swimming in shear-thinning fluids scales with a fluid's effective viscosity and can be predicted using fluid rheology and simple swimming kinematics. Our results agree reasonably well with previous theoretical predictions and provide a framework for understanding the cost of swimming in generalized Newtonian fluids.
\end{abstract}

\maketitle

\section{Introduction}
Swimming microorganisms are integral to many natural processes, including the formation of infectious biofilms in the stomach~\citep{Celli2009}, the movement of sperm cells in cervical fluid~\citep{Katz1980, Fauci2006}, and the aeration of soil by nematodes~\citep{Alexander1991}. Typical length scales for these organisms range from the micron (e.g. \textit{Escherichia coli}) to the millimetre scale (e.g. \textit{Caenorhabditis elegans}). These small length scales naturally lead to small Reynolds numbers, defined as $\mbox{\textit{Re}} = \rho U L / \eta \ll 1$, where $U$ is the swimming speed, $L$ is a characteristic length scale, and $\rho$ and $\eta$ are the fluid's density and viscosity, respectively. In this regime viscous forces dominate inertial forces, and Stokes' flow governs fluid transport, which is independent of time. Consequently, the organism's body geometry dictates the motion of a low-$\mbox{\textit{Re}}$ swimmer. Such an organism must employ a kinematically-irreversible swimming stroke for net translation to occur~\citep{Purcell1977}.

While many studies have sought an understanding self-propulsion at low $\mbox{\textit{Re}}$ in Newtonian fluids~\citep{Taylor1951, Lighthill1976, Lauga2009, Guasto2010}, there are crucial biological systems in which microorganisms must swim in complex fluids that contain polymers, particles, and large proteins~\citep{Spagnolie2015}. Examples include sperm cells in cervical mucus~\citep{Katz1980, Fauci2006} and Lyme disease spirochaetes in tissues~\citep{Harman2012}. These complex fluids typically display non-Newtonian rheological behaviour such as shear-thinning viscosity and viscoelasticity~\citep{Larson1999}. 

Recent studies of the behaviour of single swimmers in non-Newtonian fluids have centred on the effects of fluid elasticity and local structure on propulsion speed and kinematics~\citep{Lauga2007, Fu2009, Leshansky2009, Fu2010, Teran2010, Juarez2010, Shen2011, Liu2011, Harman2012, Gagnon2013, Thomases2014, Patteson2015, Qin2015}. These studies have shown that fluid elasticity modifies the swimming speed and kinematics of microorganisms. Whether swimming speed is increased or decreased seems highly dependent on the swimming gait of the organism and its coupling with the material properties of the fluid. There has been considerably less work, however, on the effects of shear rate-dependent viscosity on the swimming behaviour of microorganisms. A recent experimental investigation~\citep{Gagnon2014} found that shear-thinning viscosity does not modify the propulsion of \textit{C. elegans}, a model undulatory swimmer, when compared to Newtonian fluids of the same effective viscosity despite substantial changes in the resulting velocity fields. Previous numerical simulations and theoretical studies of swimming speed in shear-thinning fluids have predicted an enhancement or no change in an organism's swimming speed, depending on the model swimmer~\citep{Velez2013, MJ2012, MJ2013}. 

Despite recent progress, our understanding of swimming in shear-thinning fluids is still incomplete and many important questions remain, including whether a microorganism finds swimming in shear-thinning fluids mechanically easier (or more difficult) than in Newtonian fluids. One can think of this ease or difficulty in swimming as an organism's cost of swimming, or mechanical power. Our previous experiments~\citep{Gagnon2014} with swimming \textit{C. elegans} in shear-thinning fluids suggests, to first order, that the cost of swimming in an shear-thinning fluid is similar to that in a Newtonian fluid with the same \textit{effective viscosity}, which is lower than the equivalent mechanical power for a Newtonian fluid with the same \textit{zero-shear viscosity}. Recent work, both theoretical~\citep{Velez2013} and numerical~\citep{Li2015}, has also proposed a reduction in the cost of swimming for undulatory organisms in shear-thinning fluids. Indeed, it seems quite reasonable that a fluid possessing decreasing viscosity with increasing shear rate might reduce the cost of swimming, although there has been little experimental evidence for such behaviour.

\begin{figure*}
\centerline{\includegraphics[width=\textwidth]{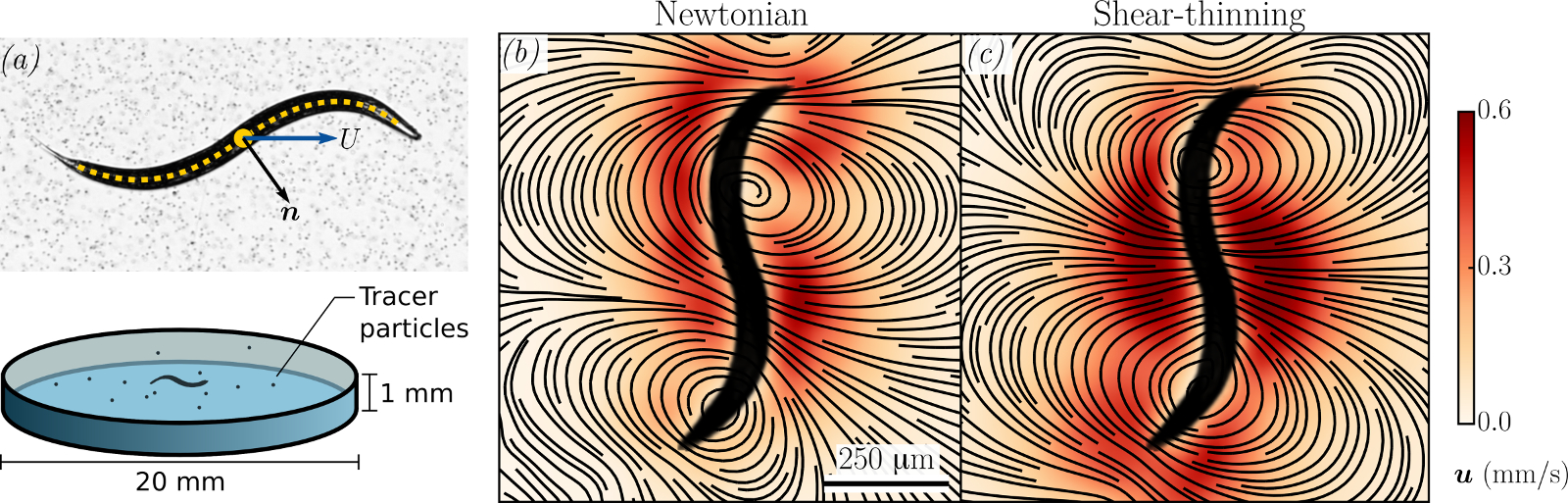}}
\caption{(Colour available online) \textit{(a)} The 1-mm nematode \textit{C. elegans} is placed in a sealed fluidic chamber containing either shear-thinning (aqueous solutions of xanthan gum) or Newtonian fluids (e.g. buffer solution). These fluids are seeded with 3.1~\si{\micro}m particles for particle tracking velocimetry. The yellow dashed line shows the tracked centreline of the nematode, with a large dot indicating its centroid. Arrows show the forward swimming speed $U$ and a sample normal vector $\boldsymbol{n}$. \textit{(b)} Snapshot of the streamlines obtained using particle tracking techniques around \textit{C. elegans} in a Newtonian buffer solution ($\eta = 1$~mPa$\cdot$s) at the moment of maximum fluid velocity. Colour represents the magnitude of the local velocity field. (c) Snapshot of the streamlines around \textit{C. elegans} during the same phase of motion in a strongly shear-thinning fluid (Carreau timescale $\lambda_{\mbox{\textit{Cr}}} \approx 6.5$ and power-law index $n \approx 0.5$). See section Fluids \& Rheology and Fig.~\ref{rheology} for more detail.}
\label{setup}
\end{figure*}

In this manuscript, we present an experimental investigation of the cost of swimming in Newtonian and shear-thinning fluids using the model biological organism \textit{C. elegans}, a 1~mm long nematode commonly used in the study of genetics and disease~\citep{Jorgensen2002, Rankin2002, Silverman2009}. We use particle tracking methods~\citep{Crocker1996, Sznitman2010PoF} to obtain the velocity fields generated by the swimming nematodes in both Newtonian and shear-thinning fluids and bright-field microscopy to track the nematode's position and surface contour during swimming~\citep{Sznitman2010BJ}. The divergence theorem \citep{Happel1983} is used to estimate the cost of swimming or mechanical power of the swimming nematode from \textit{(i)} the drag force on the nematode's body and \textit{(ii)} the viscous dissipation rate in the flow field around the swimmer. We find that the cost of swimming for weakly shear-thinning fluids are similar to Newtonian fluids of the same viscosity, whereas strongly shear-thinning fluids decrease the cost of swimming relative to Newtonian fluids. We compare our results to recent theoretical predictions~\citep{Velez2013,Li2015}.

\section{Methods}

\subsection{Experimental Techniques}
Swimming experiments in Newtonian and shear-thinning fluids are performed using the nematode \textit{C. elegans}. These nematodes are characterized by a relatively long and quasi-cylindrical body shape and are approximately 1~mm in length and 80~\si{\micro}m in diameter; their genome has been completely sequenced~\citep{Brenner1974} and a complete cell lineage has been established~\citep{Byerly1976}. \textit{C. elegans} are equipped with 95 muscle cells that are highly similar in both anatomy and molecular makeup to vertebrate skeletal muscle~\citep{White1986}. Their neuromuscular system controls their body undulations, allowing \textit{C. elegans} to swim, dig, and crawl through diverse environments. The wealth of biological knowledge accumulated to date makes \textit{C. elegans} an ideal candidate for investigations that combine aspects of biology, biomechanics, and the fluid mechanics of propulsion.

We place nematodes into sealed fluidic chambers (Fig.~\ref{setup}\textit{(a)}) that are 2~cm in diameter and 1~mm in depth, and image their swimming motion using standard bright-field microscopy (Infinity K2/SC microscope with an in-system amplifier, a CF-3 objective, and an IO Industries Flare M180 camera at 150 frames per second). The depth of focus of the objective is approximately 20~\si{\micro}m and the focal plane is set on the longitudinal axis of the nematode body. The nematode beats primarily in the observation plane; the out-of-plane beating amplitude of \textit{C. elegans} is less than 6\% of the amplitude of its in-plane motion~\citep{Sznitman2010PoF}. All data presented here pertain to nematodes swimming at the centre of the fluidic chamber and out-of-plane recordings are discarded to avoid nematode-wall interactions and to minimize three-dimensional flow effects.

Two main types of experiments comprise our study: nematode tracking and flow velocimetry. We use in-house software~\citep{Krajacic2012} to track the swimming motion of \textit{C. elegans}, extract the nematode's body-shape, and compute the nematode's kinematic properties swimming speed $U$, amplitude $A$, and frequency $f$. Our previous experiments show that \textit{C. elegans} exhibit a predominately two-dimensional sinusoidal beating pattern, producing a travelling wave that moves from head to tail~\citep{Sznitman2010PoF}. They move with a nearly constant frequency of 2~Hz and an average speed of 0.35 mm/s in water~\citep{Sznitman2010PoF}. Furthermore, recent experimental results of~\citet{Gagnon2014} indicate that the nematode's kinematics are largely insensitive to shear-thinning effects. The flow fields, however, are significantly modified by shear-thinning rheology. Please see \citet{Gagnon2014} for more details.

We measure the velocity fields generated by swimming \textit{C. elegans} in both Newtonian and shear-thinning fluids by seeding the working fluids with 3.1~\si{\micro}m polystyrene tracer particles, which are tracked continuously for the entire duration of the experiment using in-house codes. These tracer particles are dilute ($<\!0.5$\% by volume) and do not alter the properties of the fluid. We image the nematodes swimming through this seeded fluid for 6 to 10 cycles, with each swimming cycle (or period) containing 60 phases. Because \textit{C. elegans} beat at a constant frequency, we can phase-average the data and obtain spatially resolved velocity fields. We note that data points for each phase are averaged into gridded spaces of size 21~\si{\micro}m. Figure~\ref{setup}\textit{(b)} and \textit{(c)} show snapshots of streamlines computed from phase-averaged velocity fields for Newtonian and shear-thinning fluid cases, respectively; the snapshots are color-coded by the magnitude of the flow velocity. These snapshots reveal a redistribution of fluid velocity from near the head in the Newtonian case towards the tail in the shear-thinning case; additionally, shear-thinning rheology can increase both the vorticity and circulation of the body vortices produced by the nematode \citep{Gagnon2014}.

\subsection{Fluids \& Rheology}
We use both Newtonian and shear-thinning fluids in this investigation. The range of Reynolds numbers, defined as $Re=\rho UL/\eta$, is $10^{-4} < \mbox{\textit{Re}} \leq 0.35$ across all experiments, where $\mbox{\textit{Re}} = 0.35$ represents the water-like case. For shear-thinning fluids, we use $\eta$ measured at the estimated mean strain rate (from velocity fields) in computing the $Re$, which defines the fluid effective viscosity $\eta_{eff}$ in our experiments. We could alternately use the zero-shear-rate viscosity $\eta_0$ instead, but this choice would underestimate $\mbox{\textit{Re}}$. Since the non-Newtonian fluid viscosity is rate-dependent, the use of the mean strain-rate to estimate $\eta$ seems appropriate. 

\begin{figure*}
\centerline{\includegraphics[width=\textwidth]{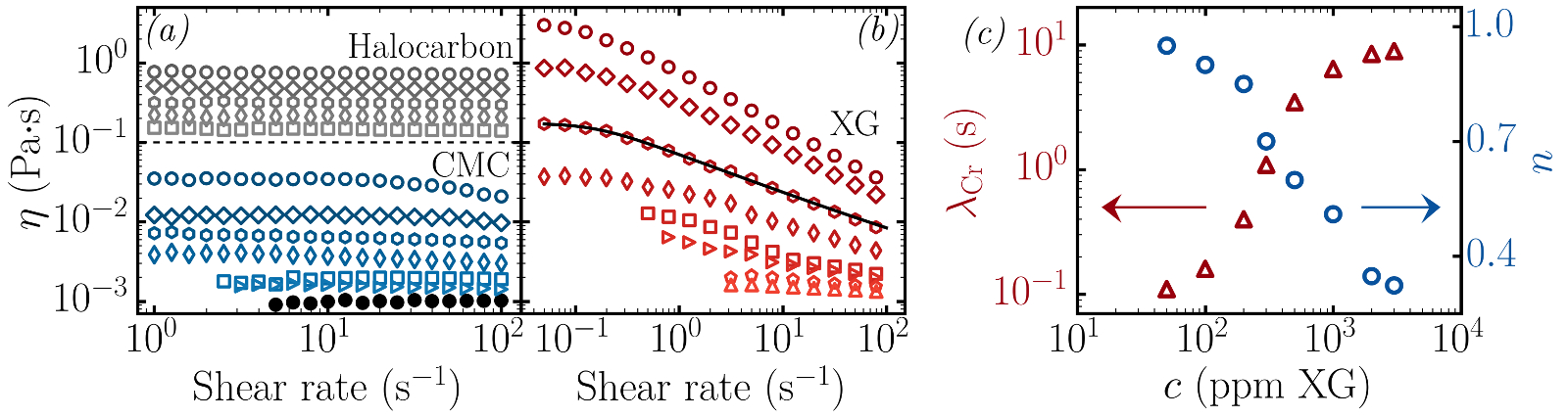}}
\caption{(Colour available online) \textit{(a)} Measurements of viscosity $\eta$ as a function of shear rate $\dot{\gamma}$ for the Newtonian buffer solution M9 (closed symbols), CMC solutions in M9 (blue open symbols, concentrations from bottom to top: 300, 500, 1000, 1500, 2000, and 3000 ppm,~\citet{Sznitman2010PoF}), and halocarbon oil mixtures (grey open symbols, from bottom to top: 100\% H27, 44\% H700, 61\% H700, 78\% H700, and 95\% H700 by weight,~\citet{Shen2011}). \textit{(b)} Measurements of viscosity $\eta$ as a function of shear rate $\dot{\gamma}$ for shear-thinning solutions of XG in M9 (from bottom to top: 50, 100, 200, 300, 500, 1000, 2000, and 3000 ppm). The solid black line shows a fit to the Carreau-Yasuda model (Eq.~\ref{Carreau}) \textit{(c)} Carreau timescale $\lambda_{\mbox{\textit{Cr}}}$ as a function of concentration ({\color{red}$\boldsymbol{\vartriangle}$}) and power law index $n$ ({\color{blue}$\boldsymbol{\circ}$}) as a function of concentration $c_{\mathrm{XG}}$.}
\label{rheology}
\end{figure*}

We prepare Newtonian fluids to cover a range of viscosities, ranging from 1~mPa$\cdot$s to 700~mPa$\cdot$s. From lowest to highest viscosity, we use \textit{(i)} a water-like buffer solution (M9 salt solution)~\citep{Brenner1974}, \textit{(ii)} very dilute solutions of the polymer carboxymethyl cellulose in M9 salt solution (CMC, $7\times10^5$ MW, Sigma Aldrich 419338), and \textit{(iii)} mixtures of two molecular weights of chlorotrifluoroethylene (halocarbon oils H27 and H700; Sigma Aldrich H8773 and H8898, respectively)~\citep{Sznitman2010PoF}. We note that halocarbon oil mixtures are limited to only Newtonian swimming kinematics data due to a significant density mismatch between these polymer solutions and polystyrene tracer particles. The polymer CMC possesses a flexible backbone, and (aqueous) solutions of CMC may exhibit viscoelasticity. Here, however, we minimize the effects of elasticity by using a low polymer concentration ($c_{\mathrm{CMC}} \lesssim 10^3$) in comparison to the overlap concentration $c_{\mathrm{CMC}}^* = 10^4$ in addition to the presence of salt (M9)~\citep{Sznitman2010PoF}. As a result, these solutions exhibit negligible shear-thinning rheology and elasticity~\citep{Shen2011}, and sample rheology curves for CMC solutions are given in Fig.~\ref{rheology}\textit{(a)}.

We prepare shear-thinning fluids by adding small amounts of the polymer xanthan gum (XG, $2.7 \times 10^{6}$ MW, Sigma Aldrich G1253) to water in the presence of salt. The XG concentration in buffer ranges from 50~ppm to 3000~ppm. These aqueous XG solutions have been well characterized and are known to have negligible elasticity~\citep{Shen2011, Gagnon2014}. We characterize all fluids (Newtonian and shear-thinning) using a cone-and-plate rheometer (strain-controlled RFS III, TA Instruments) at a range of constant shear rates. Fig.~\ref{rheology}\textit{(a)} and Fig.~\ref{rheology}\textit{(b)} show the fluid viscosity $\eta$ as a function of shear rate measurements for Newtonian and shear-thinning fluids, respectively. We find strong shear-thinning behaviour (e.g. power law viscosity) for the most concentrated XG solution ($c_{\mathrm{XG}} = 3000$~ppm), and this shear-thinning behaviour gradually decreases as the concentration of XG decreases; at the lowest concentration ($c_{\mathrm{XG}} = 100$~ppm), the behaviour of the XG solutions is Newtonian-like. We quantify this shear-thinning viscosity by fitting the rheological data with the Carreau-Yasuda model~\citep{Carreau1997}:
\begin{equation}
{\eta \left(\dot{\gamma} \right)=\eta_{\infty} + \left( \eta_0 - \eta_{\infty} \right) \left( 1 + \left( \lambda_{\mbox{\textit{Cr}}}  \dot{\gamma}  \right)^2 \right)^{\frac{n-1}{2}},}
\label{Carreau}
\end{equation}
where $\eta \left(  \dot{\gamma}  \right)$ is the fluid's shear rate-dependent viscosity, $\dot{\gamma} = \left| \boldsymbol{\dot{\gamma}} \right| \equiv \sqrt{\frac{1}{2} \left(  \boldsymbol{\dot{\gamma}}: \boldsymbol{\dot{\gamma}}\right)}$ is the magnitude of the shear rate tensor $\boldsymbol{\dot{\gamma}} \equiv \frac{1}{2} \left(\boldsymbol{\nabla} \boldsymbol{u} +  \boldsymbol{\nabla} \boldsymbol{u}^\mathsf{T} \right)$, $\eta_0$ is the zero-shear viscosity, $\eta_{\infty}$ is the infinite-shear viscosity, and $n$ is the power-law index.

The characteristic timescale $\lambda_{\mbox{\textit{Cr}}}$ represents the inverse of the shear rate at which the fluid transitions from Newtonian-like to power-law behaviour; values of $\lambda_{\mbox{\textit{Cr}}}$ for each fluid are shown in Fig.~\ref{rheology}\textit{(c)}. Larger timescales indicate that the fluid exhibits shear-thinning properties at lower shear rates~\citep{Carreau1997}. Using this timescale, we can define a non-dimensional shear rate based on the kinematics of the nematode. This kinematic Carreau number describes the strength of the shear-thinning behaviour, and we define it as $\mbox{\textit{Cr}}_k= 2 \pi \lambda_{\mbox{\textit{Cr}}} f A k$, where $f$ is the beating frequency, $A$ is the average beating amplitude, and $k$ is the wave number of the swimming nematode~\citep{Li2015}. A fluid behaves Newtonian-like when $\mbox{\textit{Cr}}_k \ll 1$ with a viscosity $\eta \approx \eta_0$. When $\mbox{\textit{Cr}}_k \gtrsim 1$, the fluid exhibits shear-thinning behaviour. The power-law index $n$, shown in Fig.~\ref{rheology}\textit{(c)}, describes the sensitivity of the fluid's viscosity to changes in shear rate.

\section{Results and discussion}
We now begin our discussion on estimating the cost of swimming or mechanical power of swimming \textit{C. elegans} from velocimetry and nematode tracking data. Here, we consider the flow of an incompressible fluid at low Reynolds number. Under these conditions, the equation of motion and the continuity equation are:

\begin{equation}
{\boldsymbol{\nabla} \cdot \boldsymbol{\sigma} = -\boldsymbol{\nabla} p + \boldsymbol{\nabla} \cdot \boldsymbol{\tau} =0}
\label{stokes}
\end{equation}
\begin{equation}
\boldsymbol{\nabla} \cdot \boldsymbol{u} = 0,
\label{continuity}
\end{equation}
where $\boldsymbol{u}$ is the fluid velocity and $\boldsymbol{\sigma}$ is the total stress tensor. The stress is defined as
\begin{equation}
{\boldsymbol{\sigma} = -p \boldsymbol{I} + \boldsymbol{\tau},}
\label{stress}
\end{equation}
where $p$ is pressure, $\boldsymbol{I}$ is the identity tensor, and $\boldsymbol{\tau}$ is the shear (deviatoric) stress. Steady flow is assumed since the ``frequency'' Reynolds number is much less than one: $Re_{f}= \rho A^2 f/\mu \ll 1$, where $f$ and $A$ are the nematode's beating frequency and amplitude, respectively~\citep{Childress1981}.

Next, we consider the energy expenditure of a swimming nematode under the above conditions and assumptions. Conservation of energy requires the power expended by the swimming nematode by deforming its body to be equal to the energy dissipation rate of the surrounding fluid. This relationship naturally provides two methods for estimating the cost of swimming or mechanical power as we will see below. We define the mechanical power associated with the motion of the nematode surface $S$ as its rate of work:
\begin{equation}
P = -\int_S \boldsymbol{n} \cdot \boldsymbol{\sigma} \cdot \boldsymbol{u} \, \mathrm{d}S,
\label{powerE}
\end{equation}
where $\boldsymbol{u}$ is the velocity of the surface. We assume that inertial and body forces are negligible for a swimming nematode, and the only forces acting on the swimmer are viscous surface forces $\boldsymbol{F} = \int_S \boldsymbol{n} \cdot \boldsymbol{\sigma} \, \mathrm{d}S = 0$.
This integral of the surface force must be zero since the swimmer is self-propelled and force-free~\citep{Lauga2009}.

Next, we can apply the divergence theorem to Eq.~\ref{powerE} to transform the surface integral into a volume integral over the surrounding fluid $V$ with the assumption that $\boldsymbol{u}$ vanishes far from the swimmer~\citep{Lighthill1976}. This transforms the surface's rate of work into the rate of viscous dissipation of the fluid:
\begin{equation}
P=-\int_S \boldsymbol{n} \cdot \boldsymbol{\sigma} \cdot \boldsymbol{u} \, \mathrm{d}S = \int_V \boldsymbol{\nabla} \cdot \left( \boldsymbol{\sigma} \cdot \boldsymbol{u} \right) \mathrm{d}V.
\label{divergence}
\end{equation}
Distributing the divergence operator on the volume integral yields:
\begin{equation}
P=-\int_S \boldsymbol{n} \cdot \boldsymbol{\sigma} \cdot \boldsymbol{u} \, \mathrm{d}S = \int_V \left( \boldsymbol{\nabla} \cdot \boldsymbol{\sigma} \right) \cdot \boldsymbol{u} + \boldsymbol{\sigma} : \boldsymbol{\nabla} \boldsymbol{u} \, \mathrm{d}V.
\end{equation}
By Stokes equation (Eq.~\ref{stokes}), the first term in the volume integral must be zero. Furthermore, we can substitute $\boldsymbol{\sigma}$ in the volume integral with the definition of the stress tensor $\boldsymbol{\sigma}$ (Eq.~\ref{stress}) so that:
\begin{equation}
P=-\int_S \boldsymbol{n} \cdot \boldsymbol{\sigma} \cdot \boldsymbol{u} \, \mathrm{d}S = \int_V -p \boldsymbol{I} : \boldsymbol{\nabla} \boldsymbol{u} + \boldsymbol{\tau} : \boldsymbol{\nabla} \boldsymbol{u} \, \mathrm{d}V.
\end{equation}
Note that $\boldsymbol{I} : \boldsymbol{\nabla} \boldsymbol{u} = \boldsymbol{\nabla} \cdot \boldsymbol{u} =0$, for an incompressible fluid, yielding the energy balance:
\begin{equation}
P= -\int_S \boldsymbol{n} \cdot \boldsymbol{\sigma} \cdot \boldsymbol{u} \, \mathrm{d}S = \int_V \boldsymbol{\tau} : \boldsymbol{\nabla} \boldsymbol{u} \, \mathrm{d}V.
\label{energyBalance}
\end{equation}
Lastly, the right-hand side of Eq.~\ref{energyBalance} is equal to the fluid's viscous dissipation:
\begin{equation}
\Phi = \int_V \boldsymbol{\tau} : \boldsymbol{\nabla} \boldsymbol{u} \, \mathrm{d} V.
\label{dissipation}
\end{equation}

Equation~\ref{energyBalance} reveals two methods for estimating the cost of swimming via both a calculation of the swimmer's mechanical power ($P$, Eq.~\ref{powerE}) and an estimate of the surrounding fluid's viscous dissipation rate ($\Phi$, Eq.~\ref{dissipation}).  In what follows, we will use Equation~\ref{energyBalance} along with experimental data (nematode tracking, velocimetry, and rheology) to estimate the cost of swimming for \textit{C. elegans}. There are three necessary ingredients: \textit{(i)} the instantaneous position of the surface $S$ (obtained from nematode tracking) and the corresponding fluid volume $V$, \textit{(ii)} a spatially differentiable flow field $\boldsymbol{u}$ (from particle tracking), and \textit{(iii)} a constitutive model for the fluid stresses $\boldsymbol{\sigma}$ (from rheology and Equation~\ref{Carreau}). First, we measure the instantaneous position of the surface $S$ and its outward normal $\boldsymbol{n}$ by tracking the body of the nematode using bright-field microscopy (Fig.~\ref{setup}\textit{(a)}). Image processing provides an outline of the nematode's body in a two-dimensional plane; to estimate $S$, we multiply the observed body shapes by the diameter of the nematode's body (80~\si{\micro}m) to form a thin surface area. For our estimate of the viscous dissipation rate, we consider the area formed by the plane of observation in our region of interest (approximately 2~mm by 2~mm). Beyond this region of interest, the velocities of the fluid are below the noise level of our particle tracking measurements. Similar to our surface integral, we multiply this area by the diameter of the nematode's body (80~\si{\micro}m) to form a small, thin volume. For both integrals (surface and volume), we note this assumes a uniform planar flow field within 40~\si{\micro}m of the mid-plane of the worm. The thinness of the selected volume aims to minimize errors associated with the three-dimensional nature of the flow. Second, using particle tracking velocimetry, we measure the velocity fields $\boldsymbol{u}$ and, because $\boldsymbol{u}$ is spatially resolved, the associated shear rate tensor $\boldsymbol{\dot{\gamma}}$ (Fig.~\ref{setup}\textit{(b)} and \textit{(c)}). 

Lastly, we consider the constitutive equation necessary to estimate the fluid stress tensor $\boldsymbol{\sigma} = -p \boldsymbol{I} + \boldsymbol{\tau}$. The pressure $p$ is found by integrating Stokes equation from the boundary of our velocity field to the fluid-swimmer interface; note that the contribution of the isotropic static pressure $p_0$ vanishes over a closed integral and as a result is negligible. For Newtonian fluids, we simply use Newton's law of viscosity $\boldsymbol{\tau} = \eta \boldsymbol{\dot{\gamma}}$. For shear-thinning fluids, the shear (deviatoric) stress is estimated for each beating phase by first differentiating the velocity field to obtain a shear rate field. We then use the Carreau-Yasuda model (Eq.~\ref{Carreau}) along with the rheological data shown in Fig.~\ref{rheology}\textit{(b,c)} to calculate a viscosity field for all shear-thinning fluids.

\begin{figure*}
\centerline{\includegraphics[width=\textwidth]{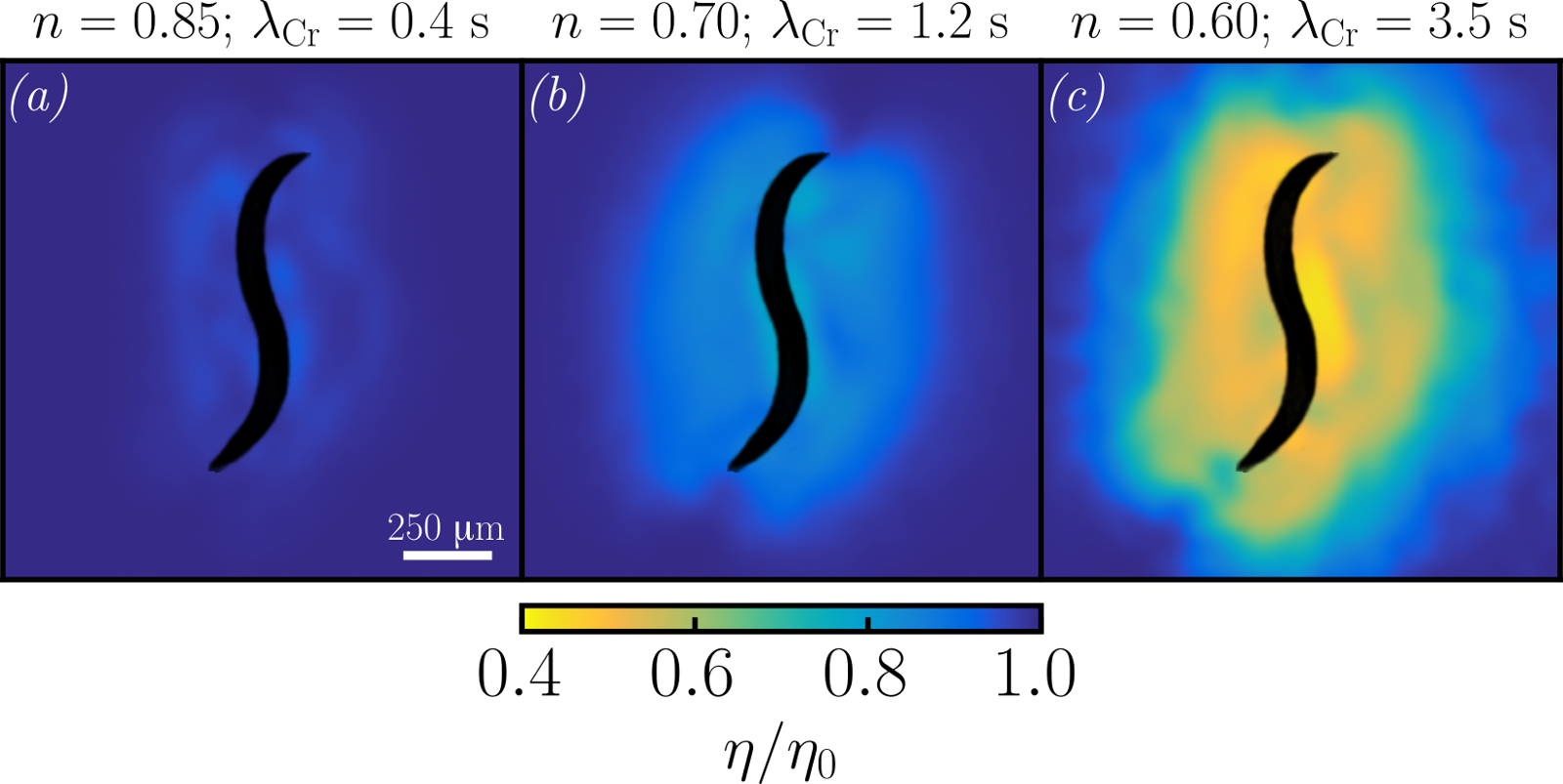}}
\caption{(Colour available online) Viscosity fields, normalized by the fluid's zero-shear viscosity $\eta_0$, for a selection of XG solutions at an instant in time: \textit{(a)} 200 ppm, \textit{(b)} 300 ppm, \textit{(c)} 500 ppm XG.}
\label{viscosity}
\end{figure*}

Figure~\ref{viscosity} shows snapshots of the spatial viscosity fields normalized by zero-shear viscosity $\eta/\eta_0$ for a particular beating phase in different XG solutions. Viscosity is colour-coded such that blue corresponds to zero-shear viscosity and yellow highlights regions of strong shear-thinning behaviour; one can also consider locations with decreased viscosity as highlighting regions of large shear rate magnitude in the flow. Figure~\ref{viscosity}\textit{(a)} shows the estimated viscosity field for the 200 ppm XG solution, which has a power-law index of $n=0.85$ and a Carreau timescale of $\lambda_{\mbox{\textit{Cr}}}=0.4$~s. The largest decrease in normalized viscosity for this fluid is approximately 15\%.  Figure~\ref{viscosity}\textit{(b)} shows the result for a solution of $n=0.70$ and $\lambda_{\mbox{\textit{Cr}}}=1.2$~s, which exhibits a decrease in viscosity of nearly 30\%. Finally, Fig.~\ref{viscosity}\textit{(c)} shows the viscosity fields for nematode's swimming in a XG solution of $n=0.60$ and $\lambda_{\mbox{\textit{Cr}}}=3.5$~s. Here, we observe the formation of a highly-thinned fluid envelope around the nematode that extends approximately 0.5 mm (or half body length) away from the nematode's body; the viscosity decrease near the swimmer is approximately 60\%. The viscosity fields for nematodes swimming in highly shear-thinning fluids, $n<0.5$ and $\lambda_{\mbox{\textit{Cr}}}>5$ (not shown), show similar behaviour to Fig.~\ref{viscosity}\textit{(c)}, and the normalized viscosity can decrease by more than an order of magnitude (from 1 to 0.1).

Measurements of the surface $S$, differentiable velocity fields $\boldsymbol{u}$, and the stress $\boldsymbol{\sigma}$ allow us to investigate whether a local decrease in viscosity modifies the cost of swimming. We compute both the mechanical power of the swimmer ($P$, Eq.~\ref{powerE}) and the rate of viscous dissipation in the fluid ($\Phi$, Eq.~\ref{dissipation}). This allows for a interesting comparison between different methods of computing the mechanical power of swimming organisms at low $\mbox{\textit{Re}}$ from experimental data. We note that particle tracking techniques can only resolve velocity fields and shear rates 40~\si{\micro}m from the boundary of the swimmer. Our calculation of drag force and power (Eq.~\ref{powerE}) are therefore somewhat hindered by our ability to make measurements close to the swimmer. We then compare our estimations of power from experimental data to recent theoretical~\citep{Velez2013} and numerical results~\citep{Li2015}.

\begin{figure*}
\centerline{\includegraphics [width=\textwidth] {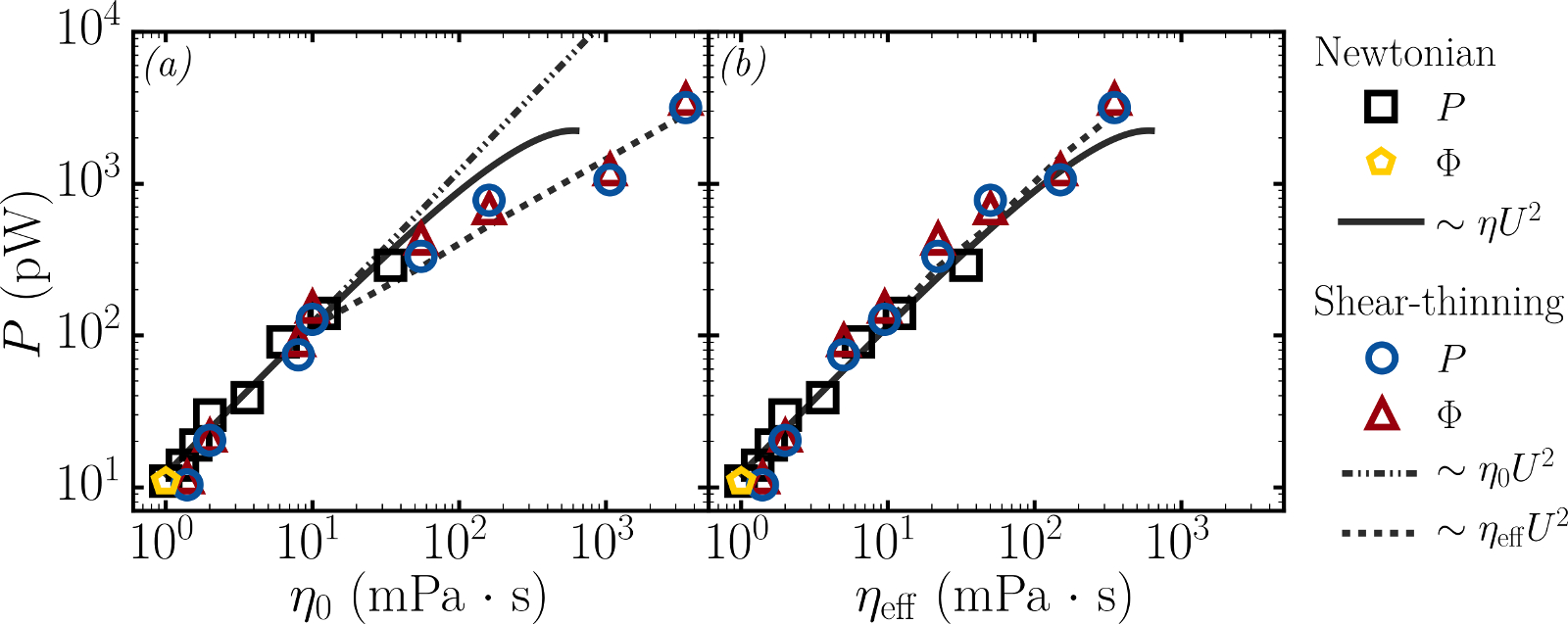}}
\caption{(Colour available online) \textit{(a)} Cost of swimming as a function of zero-shear viscosity $\eta_0$ using each side of Eq.~\ref{energyBalance}. For Newtonian fluids: mechanical power (${\square}$, from \citet{Sznitman2010PoF}), viscous dissipation rate ({\color{yellow}$\boldsymbol{\pentagon}$}, buffer only), and the scaling $P \sim \eta U^2$ (solid line) calculated from our kinematics data~\citep{Gagnon2014}. For shear-thinning fluids: mechanical power ({\color{blue}$\boldsymbol{\circ }$}), viscous dissipation rate ({\color{red}${\vartriangle}$}), and the scalings $P \sim \eta_0 U^2$ (dash-dot line) and $P \sim \eta_{\mathrm{eff}} U^2$ (dashed line). \textit{(b)} Mechanical power and viscous dissipation rate replotted versus effective viscosity $\eta_{\mathrm{eff}}$.}
\label{power}
\end{figure*}

Results for both mechanical power and viscous dissipation rate are shown in Fig.~\ref{power}. The calculation of mechanical power $P$ (Eq.~\ref{powerE}) as a function of zero-shear-viscosity $\eta_{0}$ is shown for several shear-thinning fluids ({\color{blue}$\boldsymbol{\circ }$}) and Newtonian solutions (${\square}$). Also shown is the viscous dissipation rate $\Phi$ (Eq.~\ref{dissipation}) for the same shear-thinning fluids ({\color{red}${\vartriangle}$}) as well as a Newtonian buffer solution ({\color{yellow}$\boldsymbol{\pentagon}$}). Results from the measurement of mechanical power and viscous dissipation rate show quite reasonable agreement, and suggest that both methods can be used to estimate the cost of swimming of low-${\mbox{\textit{Re}}}$ organisms. 

The data presented in Fig.~\ref{power}\textit{(a)} show that the estimated mechanical power increases linearly with fluids viscosity for \textit{C. elegans} swimming in Newtonian fluids. A deviation from this linear behaviour is found for high viscosity fluids because the nematode is power-limited~\citep{Shen2011}; this deviation starts for fluid viscosities greater than approximately 30~mPa$\cdot$s. For nematodes swimming in Newtonian fluids, we expect the mechanical power to scale as $P \sim \eta U^2$, where $U$ is the swimming speed of the nematode; the black line shows this scaling using our previously obtained experimentally-measured kinematics (in Newtonian fluids) to provide values of $U$ for a range of fluid viscosities \citep{Shen2011,Gagnon2014}. This scaling indicates that power should increase linearly with viscosity and indeed the calculation of mechanical power in Newtonian fluids supports this linearity. This indicates that the nematode's kinematics (i.e. swimming speed) are largely insensitive to changes in viscosity for  $\eta < 30$~Pa$\cdot$s ~\citep{Sznitman2010PoF}. For larger values of $\eta$, however, our data show a deviation from a linear scaling for $P \sim \eta U^2$ because the nematode is power-limited. 

For shear-thinning fluids, the mechanical power at low viscosity also increases linearly with increasing zero-shear viscosity. At moderate to large viscosities ($\gtrsim 30$~mPa$\cdot$s), mechanical power increases sub-linearly with zero-shear viscosity ($\eta_0$). In order to interpret these data, Fig.~\ref{power}\textit{(a)} also shows two scalings for shear-thinning fluids generated from our experimentally-measured kinematics in shear-thinning fluids. The first scaling ($P \sim \eta_0 U^2$), shown as a dash-dot line. This curve is essentially a continuation of the linear scaling observed at low viscosities in Newtonian fluids. The second scaling ($P \sim \eta_\mathrm{eff} U^2$), shown as a dashed line, appears to captures the sub-linearity of our mechanical power calculations; the value of $\eta_\mathrm{eff}$ is defined as the average viscosity over the range of characteristic shear rates produced by the organism~\citep{Gagnon2014}. Figure~\ref{power}\textit{(b)} evaluates the robustness of this scaling; here, the mechanical power and viscous dissipation measurements are shown versus effective viscosity ($\eta_\mathrm{eff}$) and compared with the Newtonian scaling $P \sim \eta U^2$, which remains unchanged. The data collapses onto the Newtonian scaling, suggesting that an organism's cost of swimming in a generalized Newtonian fluid is reasonably well-predicted by estimating a fluid's effective viscosity for a given swimming gait. This confirms a previous hypothesis made from the kinematics of \textit{C. elegans} as a function of effective viscosity~\citep{Gagnon2014}.

Previous work on undulatory swimming in shear-thinning fluids, however, generally compares mechanical power to the equivalent Newtonian power, defined as $P_N = P(\eta_0)$, and broadly predicts that shear-thinning viscosity decreases the cost of swimming \citep{Velez2013, Li2015}. Since a shear-thinning fluid's effective viscosity must always be less than or equal to its zero-shear viscosity, our findings in Fig.~\ref{power}\textit{(a)} suggest agreement. However, these theoretical \citep{Velez2013} and numerical \citep{Li2015} studies also suggest theoretical scalings using the Carreau-Yasuda model (Eq.~\ref{Carreau}) for mechanical power relative to the Newtonian case $P/P_N$. While the work of \citet{Velez2013} uses a small-amplitude approximation and diverges at high Carreau number $\mbox{\textit{Cr}}_k$, \citet{Li2015} have recently extended this scaling to large amplitudes:
\begin{equation}
P/P_{N} = 1 - \left(1-\eta_{\infty}/\eta_0\right)\left(1-\left(1+\mbox{\textit{Cr}}_k^2\right)^
{(n-1)/2}\right).
\label{prediction}
\end{equation}
We note that \citet{Li2015} multiply the square of the kinematic Carreau number $ \mbox{\textit{Cr}}_k$ by the constant 3/8, such that the first term of the Taylor expansion of Eq.~\ref{prediction} matches the theoretical power of small amplitude undulatory sheet in a generalized Newtonian fluid~\citep{Velez2013, Li2015}. Since we are not performing an expansion of this equation and are instead inserting experimentally-measured swimming kinematics and fluid rheology, we take this constant to be one in our experimental system.

\begin{figure*}
\centerline{\includegraphics [width=0.55\textwidth] {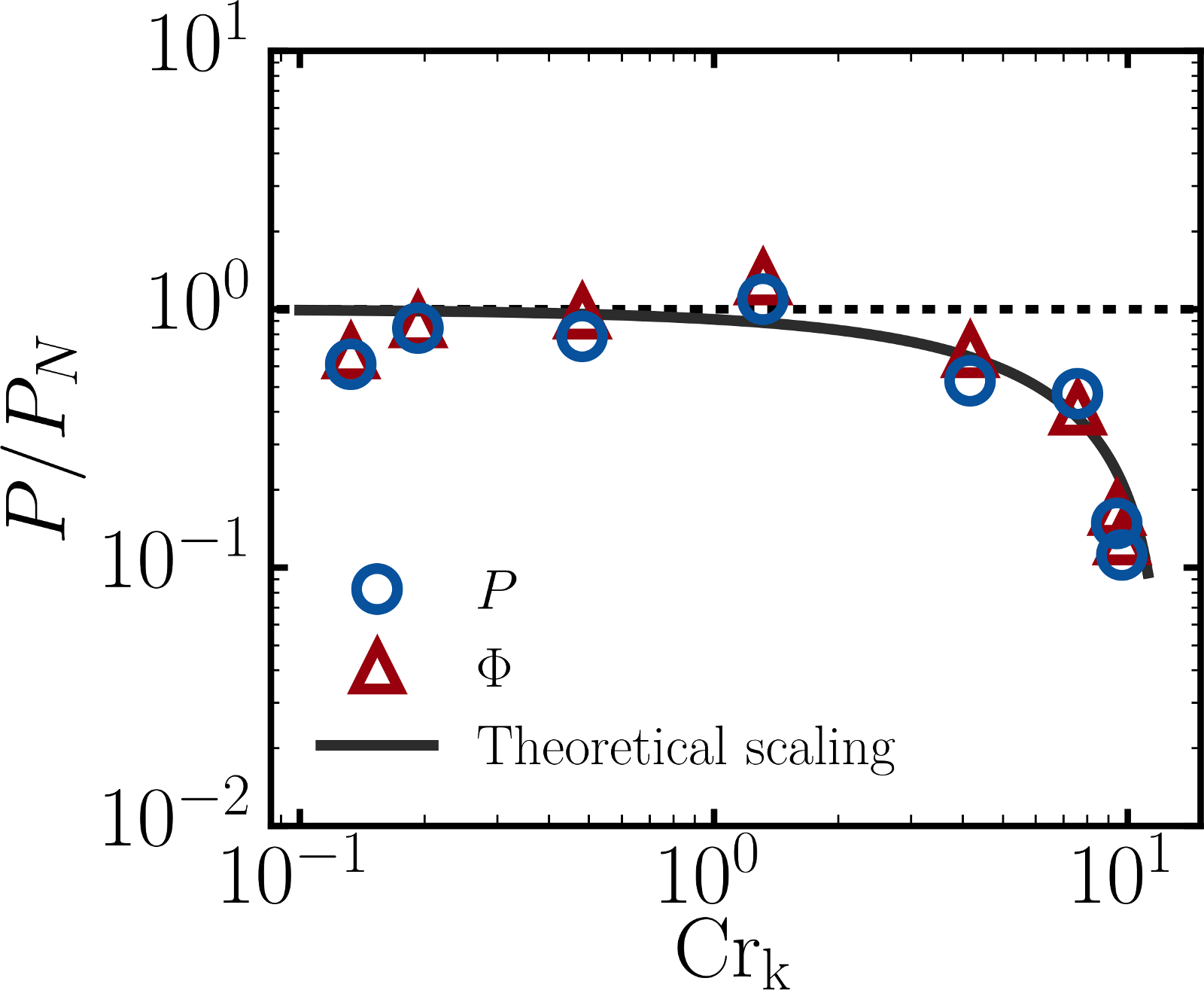}}
\caption{(Colour available online) Normalized mechanical power ({\color{blue}$\boldsymbol{\circ}$}) and viscous dissipation rate ({\color{red}$\boldsymbol{\vartriangle}$}) as a function of $\mbox{\textit{Cr}}_k$; the dashed line represents the Newtonian case. The transition from $P/P_N \approx 1$ to $P/P_N < 1$ occurs at $\mbox{\textit{Cr}}_k = \mathcal{O}(1)$. The solid black line is the theoretical scaling generated from our rheology and kinematics data, given by Eq.~\eqref{prediction}~\citep{Li2015}.}
\label{scaling}
\end{figure*}

This theoretical scaling for relative power (Eq. 3.10) now gives us a method for making a quantitative comparison between the proposed scaling and the methods for estimating mechanical power from Eq.~\ref{energyBalance}. By using our rheology data (Fig.~\ref{rheology}) and kinematics data \citep{Gagnon2014}, we can estimate $\eta_{\infty}/\eta_0$ and directly compute $\mbox{\textit{Cr}}_k$ to obtain an approximation of normalized power $P/P_N$. We show this experimental estimate of $P/P_N$ as a function of $\mbox{\textit{Cr}}_k$ in Fig.~\ref{scaling} alongside our measurements of mechanical power using both methods discussed above, \textit{(i)} mechanical power ($P$, {\color{blue}$\boldsymbol{\circ}$}) and \textit{(ii)} viscous dissipation rate ($\Phi$, {\color{red}$\boldsymbol{\vartriangle}$}). We note that we do not observe a strongly power-limited regime for \textit{C. elegans} swimming in shear-thinning fluids, though we observe some deviations from a linear scaling for Newtonian fluids; it is in such regime that one would expect significant deviations from the theoretical calculations, which assume the swimmer to have infinite power.

We find good agreement among between our calculations and the theoretical scaling based on our kinematics and rheology data: at $\mbox{\textit{Cr}}_k \sim \mathcal{O}(1)$, the cost of swimming in shear-thinning fluids transitions from $P/P_N \approx 1$ to $P/P_N < 1$. Indeed, we now gain considerable confidence in predicting the cost of swimming using only rheology ($\eta_0$, $\eta_{\infty}$, and $\lambda_{\mbox{\textit{Cr}}}$) and simple kinematics ($A$, $f$, and $k$). 

It follows that we can hypothesize the cost of transport for an undulatory swimmer in a generalized Newtonian fluid. For example, in a shear-thickening fluid with the same constitutive model (Eq.~\ref{Carreau}) but now with a power law index $n > 1$ and $\eta_0 < \eta_{\infty}$, we predict that a nematode would require more power compared to a Newtonian fluid of the same zero-shear viscosity. Additionally, when scaled by effective viscosity, we expect the cost of swimming in a shear-thickening fluid to collapse onto the Newtonian scaling, similar to our shear-thinning data in Fig.~\ref{power}\textit{(b)}.

\section{Conclusion}
Using each side of the energy balance for a low Reynolds number swimmer (Eq.~\ref{energyBalance}), we find that \textit{(i)} the mechanical power and \textit{(ii)} the viscous dissipation rate suggest that the cost of swimming for an undulatory swimming in shear-thinning fluids is smaller than the cost of swimming in a Newtonian fluid with the equivalent zero-shear viscosity. Furthermore, this cost of swimming is well-described by the scaling $P \sim \eta_{\mathrm{eff}} U^2$ (Fig.~\ref{power}\textit{(b)}). Our experimental observations show good agreement with a recent theoretical scaling \citep{Li2015} (Fig.~\ref{scaling}). These results provide a framework for understanding of the cost of swimming in generalized Newtonian fluids, which can be predicted using only the fluid's rheology and simple swimming kinematics.

\section*{Acknowledgements}
We thank T. Lamitina for \textit{C. elegans} and A.M. Ardekani, P.S. Ayyaswamy, A. Gopinath,  N.C. Keim, E. Lauga, G. Li, T. Montenegro-Johnson, and B. Qin for helpful discussions. This work was supported by NSF-CAREER (CBET)-0954084 and NSF-CBET-1437482. D.A. Gagnon was supported by an NSF Graduate Fellowship. 

\bibliographystyle{jfm}
\bibliography{shearThinning}

\begin{thebibliography}{42}
\expandafter\ifx\csname natexlab\endcsname\relax\def\natexlab#1{#1}\fi

\bibitem[Alexander(1991)]{Alexander1991}
{\sc Alexander, M.} 1991 {\em Introduction to soil microbiology\/}. Malabar,
  FL: R.E. Krieger.

\bibitem[Brenner(1974)]{Brenner1974}
{\sc Brenner, S.} 1974 The genetics of \emph{Caenorhabditis elegans}. {\em
  Genetics\/} {\bf 77}, 71--94.

\bibitem[Byerly {\em et~al.\/}(1976)Byerly, Cassada \& Russell]{Byerly1976}
{\sc Byerly, L., Cassada, R.C. \& Russell, R.L.} 1976 The life cycle of the
  nematode \textit{{C}aenorhabditis elegans}: {I}. wild-type growth and
  reproduction. {\em Dev. Biol.\/} {\bf 51}, 23--33.

\bibitem[Carreau {\em et~al.\/}(1997)Carreau, DeKee \& Chhabra]{Carreau1997}
{\sc Carreau, P.J., DeKee, D.C.R. \& Chhabra, R.P.} 1997 {\em Rheology of
  polymeric systems\/}. Munich: Hanser.

\bibitem[Celli {\em et~al.\/}(2009)Celli, Turner, Afdhal, Keates, Ghiran,
  Kelly, Ewoldt, McKinley, So, Erramilli \& Bansil]{Celli2009}
{\sc Celli, J.P., Turner, B.S., Afdhal, N.H., Keates, S., Ghiran, I, Kelly,
  C.P., Ewoldt, R.H., McKinley, G.H., So, P., Erramilli, S. \& Bansil, R.} 2009
  \textit{{H}eliobacter pylori} moves through mucus by reducing mucin
  viscoelasticity. {\em Proc. Natl. Acad. Sci. USA\/} {\bf 106}, 14321--14326.

\bibitem[Childress(1981)]{Childress1981}
{\sc Childress, S.} 1981 {\em Mechanics of Swimming and Flying\/}. Cambridge
  University Press.

\bibitem[Crocker \& Grier(1996)]{Crocker1996}
{\sc Crocker, J.C. \& Grier, D.G.} 1996 Methods of digital video microscopy for
  colloidal studies. {\em J. Colloid Interf. Sci.\/} {\bf 179}, 298--310.

\bibitem[Fauci \& Dillon(2006)]{Fauci2006}
{\sc Fauci, L.J. \& Dillon, R.} 2006 Biofluidmechanics of reproduction. {\em
  Annu. Rev. Fluid Mech.\/} {\bf 38}, 371--394.

\bibitem[Fu {\em et~al.\/}(2010)Fu, Shenoy \& Powers]{Fu2010}
{\sc Fu, H.C., Shenoy, V.B. \& Powers, T.R.} 2010 Low-{R}eynolds-number
  swimming in gels. {\em EPL\/} {\bf 91}.

\bibitem[Fu {\em et~al.\/}(2009)Fu, Wolgemuth \& Powers]{Fu2009}
{\sc Fu, H.C., Wolgemuth, C.W. \& Powers, T.R.} 2009 Swimming speeds of
  filaments in nonlinearly viscoelastic fluids. {\em Phys. Fluids\/} {\bf 21},
  033102--033110.

\bibitem[Gagnon {\em et~al.\/}(2014)Gagnon, Keim \& Arratia]{Gagnon2014}
{\sc Gagnon, D.A., Keim, N.C. \& Arratia, P.E.} 2014 Undulatory swimming in
  shear-thinning fluids: experiments with \textit{Caenorhabditis elegans}. {\em
  J. Fluid Mech.\/} {\bf 758}, R3.

\bibitem[Gagnon {\em et~al.\/}(2013)Gagnon, Shen \& Arratia]{Gagnon2013}
{\sc Gagnon, D.A., Shen, X.N. \& Arratia, P.E.} 2013 Undulatory swimming in
  fluids with polymer networks. {\em Europhys. Lett.\/} {\bf 104}, 14004.

\bibitem[Guasto {\em et~al.\/}(2010)Guasto, Johnson \& Gollub]{Guasto2010}
{\sc Guasto, J.S., Johnson, K.A \& Gollub, J.P.} 2010 Oscillatory flows induced
  by microorganisms swimming in two dimensions. {\em Phys. Rev. Lett.\/} {\bf
  105}, 168102.

\bibitem[Happel \& Brenner(1983)]{Happel1983}
{\sc Happel, J. \& Brenner, H.} 1983 {\em Low {R}eynolds number
  hydrodynamics\/}. Springer.

\bibitem[Harman {\em et~al.\/}(2012)Harman, Dunham-Ems, Caimano, Belperron,
  Bockenstedt, Fu, Radolf \& Wolgemuth]{Harman2012}
{\sc Harman, M.W., Dunham-Ems, S.M., Caimano, M.J., Belperron, A.A.,
  Bockenstedt, L.K., Fu, H.C., Radolf, J.D. \& Wolgemuth, C.W.} 2012 The
  heterogenous motility of the {L}yme disease spirochete in gelatin mimics
  dissemination through tissue. {\em Proc. Natl. Acad. Sci. USA\/} {\bf 109},
  3059--3064.

\bibitem[Jorgensen \& Mango(2002)]{Jorgensen2002}
{\sc Jorgensen, E.M. \& Mango, S.E.} 2002 The art and design of genetic
  screens: \textit{{C}aenorhabditis elegans}. {\em Nat. Rev. Genet.\/} {\bf 3},
  622--630.

\bibitem[Juarez {\em et~al.\/}(2010)Juarez, Lu, Sznitman \&
  Arratia]{Juarez2010}
{\sc Juarez, G., Lu, K., Sznitman, J. \& Arratia, P.~E.} 2010 Motility of small
  nematodes in wet granular media. {\em Europhys. Lett.\/} {\bf 92}~(4), 44002.

\bibitem[Katz \& Berger(1980)]{Katz1980}
{\sc Katz, D.F. \& Berger, S.A.} 1980 Flagellar propulsion of human sperm in
  cervical mucus. {\em Biorheology\/} {\bf 17}, 169--175.

\bibitem[Krajacic {\em et~al.\/}(2012)Krajacic, Shen, Purohit, Arratia \&
  Lamitina]{Krajacic2012}
{\sc Krajacic, P., Shen, X.N., Purohit, P.K., Arratia, P.E. \& Lamitina, T.}
  2012 Biomechanical profiling of \textit{{C}aenorhabditis elegans} motility.
  {\em Genetics\/} {\bf 191}, 1015--U1613.

\bibitem[Larson(1999)]{Larson1999}
{\sc Larson, R.G.} 1999 {\em The structure and rheology of complex fluids\/}.
  New York: Oxford University Press.

\bibitem[Lauga(2007)]{Lauga2007}
{\sc Lauga, E.} 2007 Propulsion in a viscoelastic fluid. {\em Phys. Fluids\/}
  {\bf 19}, 083104--083113.

\bibitem[Lauga \& Powers(2009)]{Lauga2009}
{\sc Lauga, E. \& Powers, T.R.} 2009 The hydrodynamics of swimming
  microorganisms. {\em Reports on Progress in Physics\/} {\bf 72}, 096601.

\bibitem[Leshansky(2009)]{Leshansky2009}
{\sc Leshansky, A.M.} 2009 Enhanced low-{R}eynolds-number propulsion in
  heterogeneous viscous environments. {\em Phys. Rev. E\/} {\bf 80}, 051911.

\bibitem[Li \& Ardekani(2015)]{Li2015}
{\sc Li, G. \& Ardekani, A.M} 2015 Undulatory swimming in non-newtonian fluids.
  {\em J. Fluid Mech.\/} {\bf 784}, R4.

\bibitem[Lighthill(1976)]{Lighthill1976}
{\sc Lighthill, J.} 1976 Flagellar hydrodynamics. {\em SIAM Rev.\/} {\bf 18},
  161--230.

\bibitem[Liu {\em et~al.\/}(2011)Liu, Powers \& Breuer]{Liu2011}
{\sc Liu, B., Powers, T.R. \& Breuer, K.S.} 2011 Force-free swimming of a model
  helical flagellum in viscoelastic fluids. {\em Proc. Natl. Acad. Sci. USA\/}
  {\bf 108}, 19516--19520.

\bibitem[Montenegro-Johnson {\em et~al.\/}(2012)Montenegro-Johnson, Smith,
  Smith, Loghin \& Blake]{MJ2012}
{\sc Montenegro-Johnson, T.D., Smith, A.A., Smith, D.J., Loghin, D. \& Blake,
  J.R.} 2012 Modelling the fluid mechanics of cilia and flagella in
  reproduction and development. {\em Eur. Phys. J. E\/} {\bf 35}, 111.

\bibitem[Montenegro-Johnson {\em et~al.\/}(2013)Montenegro-Johnson, Smith \&
  Loghin]{MJ2013}
{\sc Montenegro-Johnson, T.D., Smith, D.J. \& Loghin, D.} 2013 Physics of
  rheologically enhanced propulsion: {D}ifferent strokes in generalized
  {S}tokes. {\em Phys. Fluids\/} {\bf 25}, 081903.

\bibitem[Patteson {\em et~al.\/}(2015)Patteson, Gopinath, Goulian \&
  Arratia]{Patteson2015}
{\sc Patteson, A.E., Gopinath, A., Goulian, M. \& Arratia, P.E.} 2015 Running
  and tumbling with \textit{{E}. coli} in polymeric solutions. {\em Sci.
  Rep.\/} {\bf 5}, 15761.

\bibitem[Purcell(1977)]{Purcell1977}
{\sc Purcell, E.M.} 1977 Life at low {R}eynolds number. {\em Am. J. Phys.\/}
  {\bf 45}~(1), 3--11.

\bibitem[Qin {\em et~al.\/}(2015)Qin, Gopinath, Yang, Gollub \&
  Arratia]{Qin2015}
{\sc Qin, B., Gopinath, A., Yang, J., Gollub, J.P. \& Arratia, P.E.} 2015
  Flagellar kinematics and swimming of algal cells in viscoelastic fluids. {\em
  Sci. Rep.\/} {\bf 5}, 9190.

\bibitem[Rankin(2002)]{Rankin2002}
{\sc Rankin, C.H.} 2002 From gene to identified neuron behavior in
  \textit{{C}aenorhabditis elegans}. {\em Nat. Rev. Genet.\/} {\bf 3},
  622--630.

\bibitem[Shen \& Arratia(2011)]{Shen2011}
{\sc Shen, X.N. \& Arratia, P.E.} 2011 Undulatory swimming in viscoelastic
  fluids. {\em Phys. Rev. Lett.\/} {\bf 106}, 208101.

\bibitem[Silverman {\em et~al.\/}(2009)Silverman, Luke, Bhatia, Long, Vetica,
  Perlmutter \& Pak]{Silverman2009}
{\sc Silverman, G.A., Luke, C.J., Bhatia, S.R., Long, O.S., Vetica, A.C.,
  Perlmutter, D.H. \& Pak, S.C.} 2009 Modeling molecular and cellular aspects
  of human disease using the nematode \textit{Caenorhabditis elegans}. {\em
  Pediatr. Res.\/} {\bf 65}, 10--18.

\bibitem[Spagnolie(2015)]{Spagnolie2015}
{\sc Spagnolie, S.E.}, ed. 2015 {\em Complex Fluids in Biological Systems\/}.
  Springer.

\bibitem[Sznitman {\em et~al.\/}(2010{\natexlab{{\em a\/}}})Sznitman, Purohit,
  Krajacic, Lamitina \& Arratia]{Sznitman2010BJ}
{\sc Sznitman, J., Purohit, P.K., Krajacic, P., Lamitina, T. \& Arratia, P.E.}
  2010{\natexlab{{\em a\/}}} Material properties of \textit{{C}aenorhabditis
  elegans} swimming at low {R}eynolds number. {\em Biophys. J.\/} {\bf 98},
  617--626.

\bibitem[Sznitman {\em et~al.\/}(2010{\natexlab{{\em b\/}}})Sznitman, Shen,
  Sznitman \& Arratia]{Sznitman2010PoF}
{\sc Sznitman, J., Shen, X.N., Sznitman, R. \& Arratia, P.E.}
  2010{\natexlab{{\em b\/}}} Propulsive force measurements and flow behavior of
  undulatory swimmers at low {R}eynolds number. {\em Phys. Fluids\/} {\bf 22},
  121901.

\bibitem[Taylor(1951)]{Taylor1951}
{\sc Taylor, G.I.} 1951 Analysis of the swimming of microscopic organisms. {\em
  Proc. R. Soc. Lon. Ser.-A\/} {\bf 209}~(1099), 447--461.

\bibitem[Teran {\em et~al.\/}(2010)Teran, Fauci \& Shelley]{Teran2010}
{\sc Teran, J., Fauci, L. \& Shelley, M.} 2010 Viscoelastic fluid response can
  increase the speed and efficiency of a free swimmer. {\em Phys. Rev. Lett.\/}
  {\bf 104}, 038101.

\bibitem[Thomases \& Guy(2014)]{Thomases2014}
{\sc Thomases, B. \& Guy, R.D.} 2014 Mechanisms of elastic enhancement and
  hindrance for finite-length undulatory swimmers in viscoelastic fluids. {\em
  Phys. Rev. Lett.\/} {\bf 113}, 098102.

\bibitem[V\'{e}lez-Cordero \& Lauga(2013)]{Velez2013}
{\sc V\'{e}lez-Cordero, J.N. \& Lauga, E.} 2013 Waving transport and propulsion
  in a generalized {N}ewtonian fluid. {\em J. Non-Newton. Fluid.\/} {\bf 199},
  37--50.

\bibitem[White {\em et~al.\/}(1986)White, Southgate, Thomson \&
  Brenner]{White1986}
{\sc White, J.G., Southgate, E., Thomson, J.N. \& Brenner, S.} 1986 The
  structure of the nervous system of the nematode \textit{{C}aenorhabditis
  elegans}. {\em Phil. Trans. R. Soc. B\/} {\bf 314}, 1--340.

\end{thebibliography}

\end{document}